\documentclass[prd,aps,10pt,showkeys,nofootinbib,showpacs,twocolumn]{revtex4-1}

\usepackage{amsmath,amssymb,amsfonts,amsthm}

\usepackage{slashed}
\usepackage{mathtools}
\usepackage{amsfonts}
\usepackage{amssymb}
\usepackage{epsfig}
\usepackage{rotating}
\usepackage{url}
\usepackage{times}
\usepackage{color}
\usepackage{bm}
\usepackage{xcolor,colortbl}
\usepackage{hyperref}
\usepackage{enumitem}
\usepackage{fancyhdr}
\usepackage{anyfontsize}
\usepackage{wasysym}
\usepackage{empheq}

\newcommand{\be}{\begin{equation}}
\newcommand{\ee}{\end{equation}}
\newcommand{\bea}{\begin{eqnarray}}
\newcommand{\eea}{\end{eqnarray}}

\newcommand{\plk}{\mathfrak{h}}

\def\barr{\begin{array}}
\def\earr{\end{array}}

\def\ben{\begin{equation}}
\def\een{\end{equation}}
\def\bs{\begin{subequations}}
\def\es{\end{subequations}}
\def\bena{\begin{eqnarray}}
\def\eena{\end{eqnarray}}

\def\im{{\rm i}}

\def\bes{\begin{eqnarray}}
\def\ees{\end{eqnarray}}

\begin{document}

\title{
Lower bound on the cosmological constant from the classicality of the Early Universe}
\date{\today}

	\author{Niayesh Afshordi}
	\email{nafshordi@pitp.ca}
	\affiliation{Waterloo Centre for Astrophysics, University of Waterloo, Waterloo, ON, N2L 3G1, Canada}
\affiliation{Department of Physics and Astronomy, University of Waterloo, 200 University Ave W, N2L 3G1, Waterloo, Canada}
\affiliation{Perimeter Institute For Theoretical Physics, 31 Caroline St N, Waterloo, Canada}

	\author{Jo\~ao Magueijo}
	\email{j.magueijo@imperial.ac.uk}
	\affiliation{Theoretical Physics Group, Blackett Laboratory, Imperial College, London, SW7 2BZ, UK}

\begin{abstract}
We use the quantum unimodular theory of gravity to relate the value of the cosmological constant, $\Lambda$, and the energy scale for the emergence of cosmological classicality. The fact that $\Lambda$ and unimodular time are complementary quantum variables implies a perennially quantum Universe should $\Lambda$ be zero (or, indeed, fixed at any value). Likewise, the smallness of $\Lambda$ puts an upper bound on its uncertainty, and so a lower bound on the unimodular clock's uncertainty or the cosmic time for the emergence of classicality. Far from being the Planck scale, classicality arises at around  $7 \times 10^{11}$ GeV for the observed $\Lambda$, and taking the region of classicality to be our Hubble volume. We confirm this argument with a direct evaluation of the wavefunction of the Universe in the connection representation for unimodular theory. Our argument is robust, with the only leeway being in the comoving volume of our cosmological classical patch, which should be bigger than that of the observed last scattering surface. Should it be taken to be the whole of a closed Universe, then the constraint depends weakly on $\Omega_k$: for $-\Omega_k < 10^{-3}$ classicality is reached at $ > 4 \times 10^{12}$ GeV. 
If it is infinite, then this energy scale is infinite, and the Universe is always classical within the minisuperspace approximation. It is a remarkable coincidence that the only way to render the Universe classical just below the Planck scale is to define the size of the classical patch as the scale of non-linearity for a red spectrum with the observed spectral index $n_s = 0.967(4)$ (about $10^{11}$ times the size of the current Hubble volume). In the context of holographic cosmology, we may interpret this size as the scale of confinement in the dual 3D quantum field theory, which may be probed (directly or indirectly) with future cosmological surveys.   
\end{abstract}

\maketitle

\section{Introduction}
It is usually asserted that the Universe becomes quantum at the Planck time, but the arguments behind this are often nothing more than flimsy dimensional analysis. A closer examination shows that the issue depends on the concrete quantum gravity theory, and even then it may hinge on non-generic details (such as the choice of state or wavefunction). In this paper, we show that this is certainly the case in quantum unimodular theory~\cite{unimod1,unimod2,unimod3,unimod,Carballo-Rubio:2022ofy}, 
where the cosmological constant and unimodular (or 4-volume) time appear as quantum complementaries, subject to an uncertainty relation. This implies a relation between the non-zero value of the cosmological constant, $\Lambda$, and the emergence of large scale classicality in the early universe. 

Within such a theory, 
if $\Lambda$ were zero (or any fixed value), then the clock uncertainty would be infinite, and the Universe would be perennially quantum. More generally, stating that $\Lambda$ is small only makes sense if the uncertainty in  $\Lambda$ is smaller than its central value. This places a lower bound on the clock's uncertainty and on the time for the emergence of classicality in a unimodular theory.  Thus, a lower bound in $\Lambda$ translates into an upper bound on the temperature at the emergence of classicality, during the cosmological radiation-dominated era, which is parametrically smaller than the Planck temperature. In this paper, we will find that for the observed values of $\Lambda$ and our comoving volume, the Universe becomes classical only for  a temperatures lower than about $10^{12}$~GeV.

The argument presented here is very generic and robust, as we show in progressively great technical detail, starting in Sections~\ref{Background} and~\ref{generic} (basic argument), and closing in Sections~\ref{dynamical} and ~\ref{robust} (refinements). Indeed, the $\Lambda$ used for defining a unimodular clock does not even need to the observed $\Lambda$, should there be radiative corrections, as we show at the end of Section~\ref{robust}. This decouples our argument from some formulations of the cosmological constant problem~\cite{weinberg,padilla} (as well as from some of the corresponding solutions~\cite{pad}, which can be formulated as additions to the basic model used here~\cite{pad1}).

The only leeway is in the volume of cosmological patch where we required classical behaviour. This must be larger than the current observable Universe, but how close to this we do not know. If the Universe were closed or finite, its classical size would provide an upper bound on how large this patch is, but it could be much smaller. Conversely, in an infinite Universe, the energy scale of classicality would be infinite if classicality were required over and infinite patch (and there would be no quantum epoch, the Universe remaining classical within the minisuperspace approximation). This is because the commutation relations between $\Lambda$ and its clock involve the inverse of the comoving volume of this patch, 

Given the need for apparent fine tuning for anything between the current Hubble scale and infinity (and so an energy scale for classicality of $10^{12}$~GeV and infinity), 
in Section~\ref{SuperHor} we make a surprising discovery: The only way to render the Universe classical at the Planck scale is to define the size of the classical patch as the scale of non-linearity for a red spectrum coinciding with the observed spectral index $n_s = 0.967(4)$. This length scale is huge but not infinite: about $10^{11}$ times the size of the current Hubble volume. We further discuss the interpretation of this finding in the context of holographic cosmology.  

Finally, Section~\ref{conclude} summarizes our results and outlines future steps, including possible observational tests for the two very distinct quantum cosmologies that emerge from our analysis.      

Throughout this paper we use natural units $\hbar=c=1$ (with some exceptions, where explicit $\hbar$ is noted for clarity) and the definition of reduced Planck length: $l_P \equiv \sqrt{8\pi G} \simeq (2.44 \times 10^{18} ~{\rm GeV})^{-1}$.
\section{Background}\label{Background}
We work within the Henneaux and Teitelboim formulation of  unimodular gravity~\cite{unimod}, where full diffeomorphism invariance is preserved, but one adds to the 
base action $S_0$ (here standard General Relativity)
an additional term:
\be\label{Utrick}
S_0\rightarrow S=
S_0- \gamma \int d^4 x \, \Lambda (\partial_\mu T^\mu)
\ee
(where $\gamma$ is an arbitrary  normalization factor inserted for later convenience). 
In this expression $T^\mu$ is a  density, and so the added term is diffeomorphism invariant whislt not requiring the use of  
the metric or the connection. 
Since $T^\mu$ does not appear in $S_0$, we have:
\be\label{ConstL}
\frac{\delta S}{\delta T ^\mu}=0\implies \partial_\mu \Lambda =-\frac{1}{\gamma}\frac{\delta S_0}{\delta T^\mu}=0,
\ee
i.e. on-shell constancy of $\Lambda$. 
The other equation of motion is:
\be\label{eom2}
\frac{\delta S}{\delta \Lambda }=0\implies \partial_\mu  T^\mu=\frac{1}{\gamma}\frac{\delta S_0}{\delta \Lambda}=-\frac{\sqrt{-g}}{8\pi G_N\gamma},
\ee
(where $G_N$ is Newton's constant), and so $T^0$ is proportional to a well-known candidate for relational time: 4-volume time~\cite{unimod,Bombelli,UnimodLee2} (a 4D generalization of the earlier Misner's 3D volume time~\cite{misner}). 
Since the metric and connection do not appear in the new term, the Einstein equations (and other
field equations) are left unchanged. Thus, classically nothing changes, except that $\Lambda$ becomes a constant of motion instead of a parameter in the Lagrangian. 




However, the quantum theory is radically different. Performing a 3+1 split of the new term we find that $\Lambda$ is now a variable conjugate
to the relational time $T$. Upon quantization they become duals satisfying commutation relations. If $q^A$ represents the other degrees of freedom of matter and geometry (metric or connection), the Hamiltonian constraint can either be written in terms of $\Lambda$ (resulting in the  standard Wheeler--DeWitt equation for timeless $\psi_s(q^A,\Lambda)$) or in terms of its conjugate time $T$ 
(leading to a Schr\"odinger-like equation for $\psi(q^A,T)$)~\cite{GielenMenendez,JoaoLetter,JoaoPaper}. The general solution takes the form:
\be\label{gensol0}
\psi(q^A, T)=\int d{\Lambda} {\cal A}( \Lambda) \exp{\left[-\frac{\im }{\plk } \Lambda  T  \right]}\psi_s(q^A; \Lambda).
\ee
This is only a slight generalization of Eq.70 in Smolin's groundbreaking paper~\cite{UnimodLee2}, with $q^A$ taken to be the Ashtekar connection, and $\psi_s$ the Chern-Simons state. To the best of our knowledge this is the earliest appearance of this solution in the literature.

In what follows, unless noted otherwise, we consider a cosmological mini-superspace reduction. Then the base action (before the addition of radiation and dust matter) becomes:
\be
S_0 = \frac{3V_c}{8\pi G}\int {\rm d}t \left(\dot{b}a^2- N a\left(-(b^2+k) +\frac{\Lambda a^2}{3}\right)\right)
\label{GRmatter}
\ee
where $N$ is the lapse function, $a$ is the expansion factor, $b$ is the connection variable (on-shell $b=N\dot a$), $k$ is the curvature (taken to be 1 later in the paper), and $V_c$ is the comoving volume of the spatial region under study. It is convenient to choose 
$\gamma=3/(8\pi G_N$) in (\ref{Utrick}) so that the Poisson brackets of $\Lambda$ and $T$ mimic those of $b$ and $a^2$:
\be
\{b,a^2\} = \{\Lambda ,T\}=\frac{8\pi G}{3V_c}\, .
\ee
Then, classically (on-shell) we have:
\be
\dot T=\{T,H\}=-\frac{Na^3}{3} \implies T=-\frac{1}{3}\int dt N a^3.
\ee
Quantum mechanically, we have 
commutation relations:
\begin{equation}\label{comms}
    \left[\Lambda ,T\right]=\im \plk:=\frac{8\pi\im G \hbar}{3 V_c}=\im \frac{l_P^2}{3V_c}
\end{equation}
and the general solutions (\ref{gensol0}) have reduced form in the connection representation: 
\be\label{gensol1}
\psi(b, T)=\int d{\Lambda}\; {\cal A}( \Lambda) \exp{\left[-\frac{\im}{\plk} \Lambda  T  \right]}\psi_s(b; \Lambda).
\ee

An advantage of the unimodular extension is that it suggests a natural inner product~\cite{JoaoPaper,bbounce}:
\bea\label{innalpha}
\langle\psi_1|\psi_2  \rangle  &=&\int d \Lambda \; {\cal A}_1^\star (\Lambda) {\cal A}_2(\Lambda)\,.
\eea
This is automatically conserved, so that the theory is unitary. It allows for the construction of normalizable wave packets, whereas the original fixed-$\Lambda$ solutions, just like any plane wave, are non-normalizable. It also implies a definition of probability and a measure in $b$ space (as we spell out in Section~\ref{dynamical}; see~\cite{JoaoPaper,bbounce,LMbounce}).

In closing we note that we could subject this construction to a canonical transformation $\Lambda \rightarrow \phi(\Lambda)$ and $T\rightarrow T_\phi=T/\phi'(\Lambda)$, for a generic function $\phi$. 
All such theories are classically equivalent (and equivalent to GR), but their quantum mechanics is different. Their solutions (\ref{gensol0}) are different: a Gaussian in $\Lambda$ is not a Gaussian in a generic $\phi(\Lambda)$; the frequency $\Lambda T$ is not invariant under the canonical transformation. The natural unimodular inner product (\ref{innalpha}) is also not invariant~\cite{bbounce,JoaoPaper}. Although all these quantum theories are different, for a generic $\phi$ chosen {\it within reason}, their border with the semi-classical limit is the same, as we will comment in more detail later.


\section{Generic argument}\label{generic}

\begin{figure}
    \centering
    \includegraphics[width=1.1\linewidth]{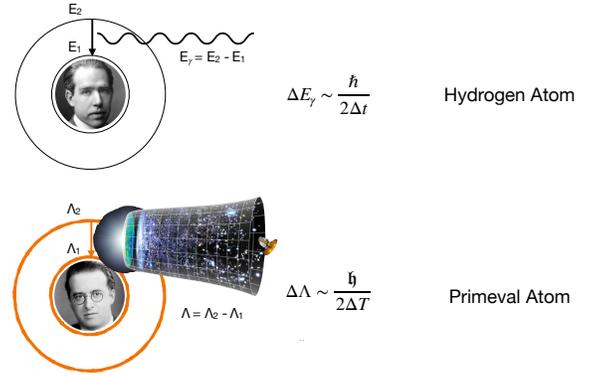}
    \includegraphics[width=\linewidth]{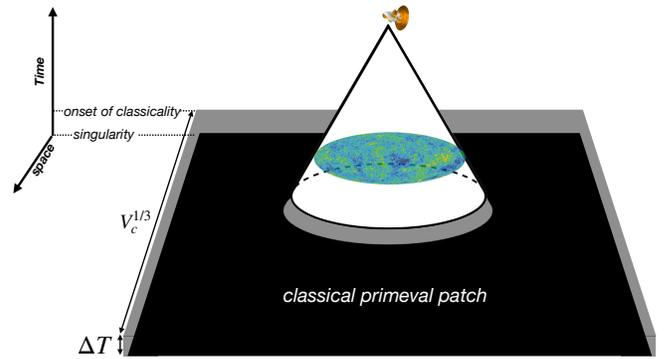}
    \caption{(Top): Analogy between the quantum creation of a photon (from Bohr's Hydrogen atom) and that of the universe (from Lema\^{i}tre's primeval atom), implying an uncertainty for the energy and cosmological constant, respectively. (Bottom): The conformal diagram of the Big Bang spacetime, indicating the past light cone, and the classical primeval patch.}
    \label{fig:atom}
\end{figure}
We first propose a generic argument that does not depend on the detailed dynamics (although it does rely on the inner product (\ref{innalpha}), and it may be argued that the inner product choice already prefigures knowledge of the dynamics~\cite{JoaoPaper,bbounce}). The fact that $\Lambda$ and $T$ satisfy commutation relations (\ref{comms}), that they are hermitian under (\ref{innalpha}), and that physical states are normalizable under this product,  implies a 
Heisenberg uncertainty relation:
\be\label{Heis1}
\sigma(\Lambda)\sigma(T)\ge \frac{\plk}{2}= \frac{l_P^2}{6V_c},
\ee
(which can be intuitively depicted in the top panel of Figure (\ref{fig:atom})).
The inequality is saturated when ${\cal A}(\Lambda)$ is a Gaussian:
\be
{\cal A}(\Lambda)=\sqrt{{\bf N}(\Lambda_{0},\sigma_\Lambda)}=
\frac{\exp{\left[-\frac{(\Lambda-\Lambda_0)^2}{4\sigma_\Lambda^2 }\right]}}
{(2\pi \sigma_\Lambda ^2)^{1/4}},\label{eq:gaussian}
\ee
so that for these states a given $\sigma_\Lambda$ translates into  the minimal $\sigma(T)=\sigma_T=\plk/(2\sigma_\Lambda)$, which, we stress, is constant in time. This is enough to derive a generic lower bound on $\Lambda$ from the fact that the Early Universe should be (semi-)classical for times $T>T_\star$, for a given time $T_\star$. In the next Section we will derive explicit solutions from the dynamics, showing that such a $\sigma_T$  translates into a $\sigma(b)$ implying the same border between classical and quantum regimes, but the generic argument in this Section may be enough for most tastes. 

A physical analogue is the broadening of the atomic emission lines (Figure \ref{fig:atom}, top panel), which is described by a Lorentzian profile:
\bea
{\cal A}(\Lambda) = \sqrt{\frac{\plk}{2\pi \sigma(T)}} \times \frac{1}{\Lambda-\Lambda_0 + \frac{i\plk}{2\sigma(T)} },\\  P(\Lambda) = \frac{\plk}{2\pi \sigma(T)}\times  \frac{1}{(\Lambda-\Lambda_0)^2+\frac{\plk^2}{4\sigma(T)^{2}}}. 
\eea
In this case, in contrast to the Gaussian wavefunction (\ref{eq:gaussian}), the variance of $\Lambda$ is divergent, but the 68\% confidence region is $\sigma^{68\%}(\Lambda) \simeq 0.9 \plk/\sigma(T)$


The main point is that in quantum unimodular theory, even when $\Lambda$ is subdominant, it supplies a quantum clock for the Universe via its conjugate\footnote{In standard unimodular theory this is the only quantum clock. In other 
theories one could consider multiple clocks at different epoch of the Universe~\cite{GielenMenendez,JoaoLetter,JoaoPaper,bbounce,LMbounce}, or even at the same epoch~\cite{twotimes}. The constraints on each of these different theories are specific to each of them.}. For the Universe to be classical at time $T$, the clock's  uncertainty $\sigma_T$ should be negligible for the relevant timing purposes, i.e.: $\sigma_T\ll T$. Hence the classicality of the Early Universe imposes a lower bound on $\Lambda$. If $\Lambda$ were zero, then $\sigma(\Lambda)=0$, implying via (\ref{Heis1}) that $\sigma_T=\infty$, and so a permanently  quantum gravitational Universe.  Generally, stating that $\Lambda$ is ``small'' only makes sense if $\sigma(\Lambda)$ is ``smaller'' than its central value $\Lambda_0$: $\sigma(\Lambda)\ll \Lambda_0$. This implies a lower bound on $\sigma_T$, and so on the time $T_\star$ when $\sigma_T\sim T$ (so that classicality occurs for $T>T_\star$).

Since we do not know how much smaller than $\Lambda_0$ the $\sigma(\Lambda)$ actually is, we 
parameterize 
$\sigma(\Lambda)=\epsilon \Lambda_0$ with $\epsilon<1$. Saturation of (\ref{Heis1}) then produces: 
\be\label{sigmaTbound}
\sigma(T)=\sigma_T=\frac{l_P^2}{6V_c\Lambda_0 \epsilon }> \frac{l_P^2}{6V_c\Lambda_0}.
\ee
This implies a Universe in the realm of quantum cosmology at times $T$ such that $\sigma(T)/T>1$, that is for $T<T_\star=\sigma_T$.
Hence we can only ignore quantum gravity at time $T$ if:
\be
T>T_\star
= \frac{l_P^2}{6V_c\Lambda_0 \epsilon }
> \frac{l_P^2}{6V_c  \Lambda_0}.
\ee
As in any quantum cosmology argument based on minisuperspace the question arises as to what $V_c$ should be. We offer 3 possibilities:
\begin{itemize}
    \item $V_c \simeq (4\pi/3)(\pi/H_0)^3$, that is the comoving volume corresponding to the present observable universe. The rationale behind this is 
    that we do not know if the Universe is classical or quantum on a larger scale. 
    \item The whole of a spherical Universe, i.e. $k=1$, $V_c=2\pi^2$ if we set $a=1$ when the Universe has unit radius. 
    If we set $a=1$ today, then the 3D volume of the Universe is $V_3 = 2\pi^2 a^3 k^{-3/2} = 2\pi^2 a^3 H^{-3}_0 (-\Omega_k)^{-3/2}$, so 
\begin{equation}\label{Vck=1}
    V_c = 2\pi^2 H^{-3}_0 (-\Omega_k)^{-3/2}.
\end{equation}
This introduces a free parameter, $\Omega_k$, in our prediction.

    \item There is an intrinsic infrared cutoff for the size of the classical primeval patch . We can parameterize 
    \begin{equation}\label{alphadef}
         V_c=\frac{\alpha}{H_0^3}
    \end{equation}
    with $\alpha>1$ varying from model to model.

\end{itemize}
We combine all of this into
\be\label{sigmaTboundb}
\sigma(T)=\sigma_T=\frac{l_P^2}{6H_0^{-3} \Lambda_0 \alpha \epsilon } = \frac{l_P^5 \sqrt{\Lambda_0}}{18 \alpha \epsilon\sqrt{3 \Omega_\Lambda^3}}
\ee
where $\alpha>1$ and $\epsilon<1$ pull in opposite directions 
and $\Omega_\Lambda \simeq 0.69$. We can also obtain (\ref{Vck=1}) by setting $\alpha=2\pi^2 (-\Omega_k)^{-3/2}$.

\begin{figure}
\includegraphics[width=\linewidth]{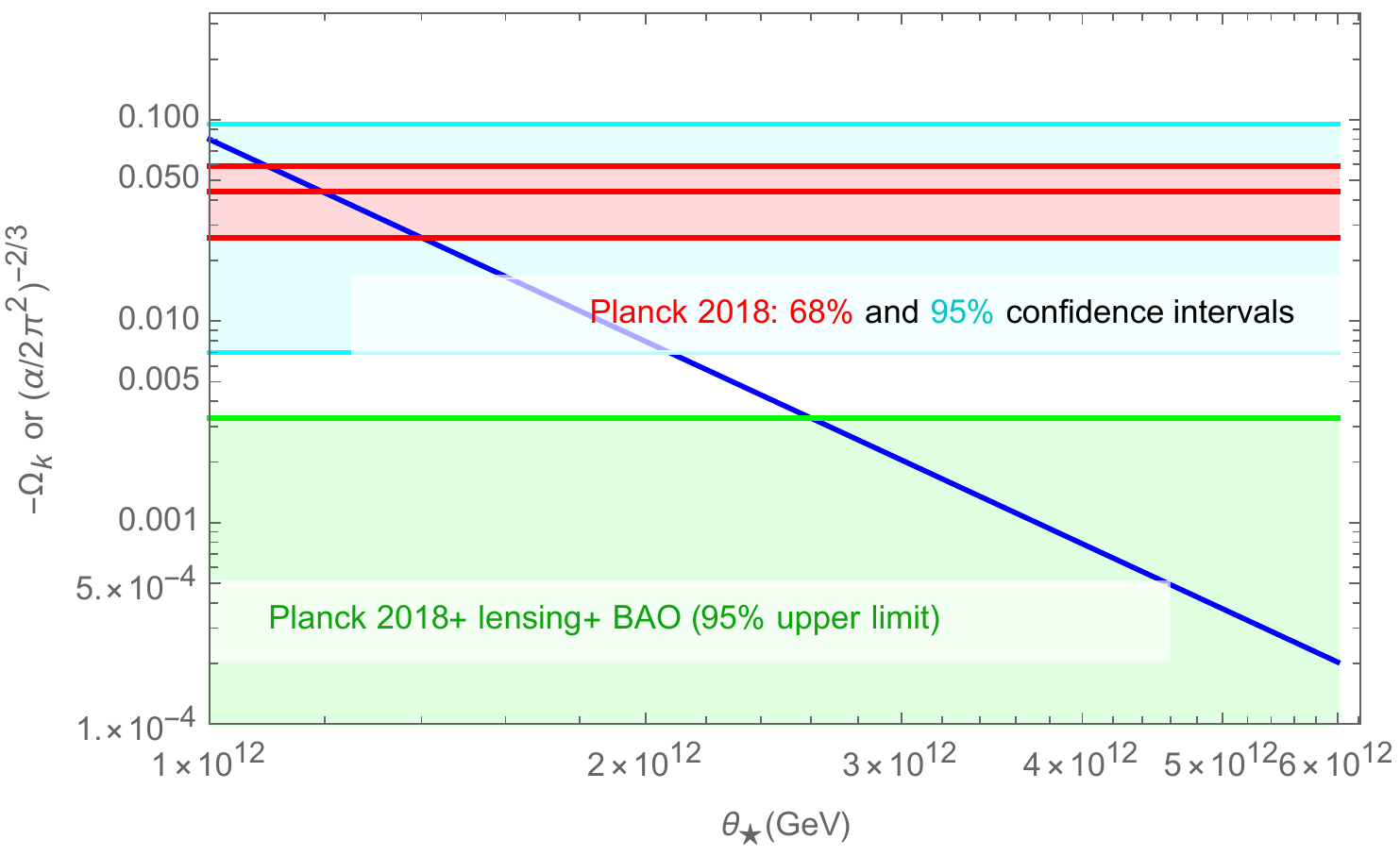}
\caption{ The relationshiop between the classicality temperature $\theta_\star$ and the spatial curvature $\Omega_k$ (Eq. \ref{t_k}), or the volume parameter of the classical primeval patch $\alpha$ (\ref{t_alpha}). Shaded areas show the 68\% and 95\% constraints on $\Omega_k$ from Planck 2018, and Planck 2018+lensing+Baryonic Acoustic Oscillations (BAO) of BOSS DR12 galaxy survey \cite{Planck:2018vyg} (which are clearly in tension, e.g., \cite{Handley:2019tkm,DiValentino:2019qzk}).}
\label{fig:omegak}
\end{figure}

We can now relate unimodular time and redshift via a change of variables:
\be\label{Tofz}
-T(z)=\frac{1}{3}\int ^\infty_z \frac{d\tilde z}{(1+\tilde z)^4 H(\tilde z)},
\ee
and use the Friedmann equation
to find:
\begin{equation}\label{Tofz1}
T(z)=-\frac{1}{15H_0\sqrt{\Omega_m}}    \frac{z^5}{\sqrt{z_{eq}}}.
\end{equation}
We can also use, under the assumption of adiabatic expansion:
\bea
H^2(z) = \frac{\pi^2}{90}l_P^2 g(z) \theta(z)^4,\\
g(z)\theta(z)^3(1+z)^{-3} = g_0 \theta^3_{\rm CMB},\\
g_0=3.91, \theta_{\rm CMB} = 2.73~{\rm K}= 9.65 \times 10^{-32} l_P^{-1}, 
\eea
to obtain:
\be
-T(z) \simeq \frac{\sqrt{10} g_0 \theta_{\rm CMB}^3}{5\pi l_P g(z)^{3/2}\theta(z)^5 } 
\ee

\be
\frac{\sigma(T)}{|T|}= \frac{5\pi l_P g(z)^{3/2}\theta(z)^5  \sigma_T}{\sqrt{10} g_0 \theta_{\rm CMB}^3}  
\ee
where we have used the fact that $\sigma(T)=\sigma_T$ does not change with time. 
Inserting (\ref{sigmaTboundb}) we arrive at:
\be
\frac{\sigma(T)}{|T|}= \frac{\sqrt{5}\pi g(z)^{3/2}}{18\alpha g_0 \sqrt{6 \Omega^3_\Lambda}} \times \frac{l_P^3 \theta^5 \sqrt{\Lambda_0}}{\theta_{\rm CMB}^3}
\ee
so that the Universe can only be classical at the relevant large scales for
\be
\theta < \theta_\star \simeq \left(  \frac{18\alpha g_0\sqrt{6 \Omega^3_\Lambda}}{\sqrt{5}\pi g_\star^{3/2}} \times \frac{\theta_{\rm CMB}^3}{l_P^3 \sqrt{\Lambda_0}} \right)^{1/5}.
\label{eq:theta_alpha}
\ee
For a whole closed Universe this becomes:
\be
\theta < \theta_\star \simeq \left(  \frac{36\pi g_0\sqrt{6 (-\Omega_\Lambda/\Omega_k)^3}}{\sqrt{5} g_\star^{3/2}} \times \frac{\theta_{\rm CMB}^3}{l_P^3 \sqrt{\Lambda_0}} \right)^{1/5}.
\ee
For $g_\star = 107$ (for ultra-relativistic Standard Model) and $\Lambda_0 \simeq 7.3 \times 10^{-121} l_P^{-2}$, these imply:
\bea
\theta_\star &\simeq& \frac{2.8 \times 10^{-7}}{l_P} \left(\alpha \over 4\pi^4/3 \right)^{1/5} \nonumber\\ &=& 6.8 \times 10^{11}~ {\rm GeV}\left(\alpha \over 4\pi^4/3 \right)^{1/5},  \label{t_alpha}
\eea
or in a closed universe:

\bea
\theta_\star &\simeq& \frac{7.7 \times 10^{-7}}{l_P} \left(-\Omega_k \over 0.01 \right)^{-3/10}  \nonumber\\  &=& 1.9 \times 10^{12}~ {\rm GeV}\left(-\Omega_k \over 0.01 \right)^{-3/10}, \label{t_k} 
\eea
which is shown in Figure (\ref{fig:omegak}), and compared to the current observational bounds on $\Omega_k$. As we see, rather than being the Planck scale, the most natural scale for the emergence of classicality in this theory is $10^{12}$~GeV. 

This is the main conclusion in this paper, and the rest of the paper is devoted to refining the argument and checking how robust it might or might not be. This conclusion will be reinforced and vindicated, until at the very end of this paper, where a surprising discovery provides an interesting exception to the rule. 




\section{Gaussian States in Mini-Superspace}\label{dynamical}
It is possible to confirm the argument in the last Section  with explicit dynamical solutions. 
For a Universe with $\Lambda$, matter and radiation, the $\psi_s$ in (\ref{gensol0}) must satisfy the Hamiltonian constraint for: 
\be\label{ham}
{\cal H}= N a\left (-(b^2+k) +  \frac{\Lambda a^2}{3}+
\frac{m_m}{a}+
\frac{m_r}{a^2}\right).
\ee
General solutions in the connection representation have been found in~\cite{JoaoPaper,bbounce,LMbounce}. They can be written as:
\be
\psi_s=\frac{1}{\sqrt{2\pi\plk}}e^{\frac{i}{\plk}P(b,\phi)}
\ee
for functions $P$ which take the simple  asymptotic forms:
\bea
P&\approx &\frac{3}{\Lambda}{\cal L}_{CS}\approx \frac{3}{\Lambda}\frac{b^3}{3}\label{CS}\\
&\approx &
C_1-\frac{2}{9}\frac{\Lambda m_m^4}{3}\frac{1}{b^9}\\
&\approx &
C_2 -\frac{1}{5}\frac{\Lambda m_r^2}{3}\frac{1}{b^5}
\eea
in the $\Lambda$, matter and radiation epochs, respectively (we have ignored the curvature $k$, and highlighted the dependence in $\Lambda$ via $3/\Lambda$). 
Here $C_1$ and $C_2$ are constants which are irrelevant for our purposes. Assuming a sharp Gaussian for ${\cal A}(\Lambda)$ we can Taylor expand $P$ and explicitly evaluate (\ref{gensol0}) as:
\bea\label{coherent0}
\psi (b,T)&=& \frac{\psi (b,T;\Lambda_0)}{(2\pi\sigma_T^2)^\frac{1}{4}}\exp\left[-\frac{(X^{\rm eff}-T )^2}{4\sigma_T^2}\right]
\eea
where
\be
\psi (b,T;\Lambda_0)=e^{ \frac{\im}{\plk}(P(b;\Lambda_0)- \Lambda_0\cdot  T)}
\ee
is the monochromatic partial wave for the central value $\Lambda=\Lambda_0$ (which is a pure phase), where:
\be
X^{\rm eff}=\frac{\partial P}{\partial \Lambda}{\Big|}_{\Lambda_{0}}\,.
\ee
and where:
\be
\sigma_T=\frac{\plk}{2\sigma_\Lambda}=  \frac{ l_P^2}{ 6V_c \sigma_\Lambda }
\ee
indeed saturates the Heisenberg bound, and therefore is constant, as assumed in the argument of the previous Section. For the 3 epochs of the Universe we have respectively:
\bea
X^{\rm eff}&\approx& -\frac{b^3}{\Lambda^2}\\
&\approx& -\frac{2}{27}\frac{m_m^4}{b^9}\\
&\approx& -\frac{1}{15}\frac{m_r^2}{b^5}.
\eea 
and it can be checked that $\dot X^{\rm eff}=\dot T=-Na^3/3$ represents the classical trajectory. This is followed by the peak of the Gaussian, so   absence of quantum behavior can be assessed from the induced
\be
\frac{\sigma(b)}{ b}\approx
\frac{\sigma(X^{\rm eff})}{b\left|\frac{\partial X^{\rm eff}}{\partial b}\right|}
=\frac{\sigma_T}{b\left|\frac{\partial X^{\rm eff}}{\partial b}\right|}
\ee
following from error propagation and $\sigma(X^{\rm eff})=\sigma_T$. Thus:
\bea
\frac{\sigma(b)}{b}&\approx&\frac{\Lambda_0^2}{3}
\frac{\sigma_T}{b^3}\\
&\approx&\frac{3}{2}\frac{\sigma_T}{m_m^4}b^9\propto b^9\propto z^{9/2}\\
&\approx&\frac{1}{3}\frac{\sigma_T}{m_r^2}b^5
\propto b^5\propto z^5
\eea
for the 3 epochs in the life of the Universe. Considering that we are just entering the Lambda epoch (so that up to factors of order 1, $H_0^2=b_0^2\sim \Lambda_0$), this implies up to factors of order 1:
\be
\frac{\sigma(b)}{b}\approx\frac{\sigma_T}{H_0} \frac{z^5}{\sqrt{z_{eq}}}\sim \frac{\sigma(T)}{|T|}
\ee
where we have used (\ref{Tofz1}) in the last step.

Hence not only do the explicit minisuperspace solutions vindicate the essential assumption that $\sigma(T)$ is a constant and that its effects translate into uncertainties in the cosmic evolution (in terms of $b$), but they do not lead to significant numerical corrections.


\section{Robustness with regards to the choice of function  $\phi(\Lambda)$}\label{robust}

The choice of $\phi(\Lambda)$ (see end of Section~\ref{Background}) leads to different quantum theories; indeed the most natural one to emerge from the dynamics is $\phi=3/\Lambda$, the ``wavenumber'' appearing in the Chern-Simons state (cf.~Eq.\ref{CS}, see also~\cite{Kodama,realkod,JoaoPaper}). However, unless $\phi$ is very contrived, this does not affect the discussion in this paper, modulo factors of order 1. 
For example, for any power-law $\phi\propto \Lambda^n$, if $\sigma(\Lambda)/\Lambda_0=\epsilon \ll 1$, then, from small error propagation:
\begin{equation}
    \frac{\sigma(\phi)}{\phi_0}\approx |n|\frac{\sigma(\Lambda)}{\Lambda_0}=|n|\epsilon
\end{equation}
and nothing changes qualitatively in our arguments unless
$|n|\gg 1$ or $|n|\ll 1$ (i.e. making the results applicable to the topical case $\phi=3/\Lambda$). More generally, if a Gaussian is sufficiently sharply peaked, then the distribution of any function of its random variable is also {\it approximately} a sharply peaked Gaussian with variance obtained by small error propagation
$\sigma(\phi)\approx \phi'\sigma(\Lambda)$.

So all we need is for $\phi'\Lambda/\phi$ to be order 1 at $\Lambda_0$.  The arguments in Sections~\ref{generic} and~\ref{dynamical} then follow. A coherent state in $\Lambda$ is  quasi-coherent in any $\phi$ and vice versa. The arguments in Section~\ref{generic} follow through because:
\begin{equation}
    \frac{\sigma(T_\phi)}{T_\phi}
    =\frac{\plk}{2\sigma(\phi)T_\phi}
    \approx 
    \frac{\plk}{2\sigma(\Lambda)T}=
    \frac{\sigma(T)}{T}.
\end{equation}
Likewise for the arguments in Section~\ref{dynamical} since:
\begin{equation*}
    X^{\rm eff}_\phi=\frac{X^{\rm eff}}{\phi'_0}
\end{equation*}
so that the extra factor in $T_\phi=T/\phi'$ cancels throughout (in the peak trajectory and in $\delta b/b$). 
Obviously we can design functions for which the argument fails because the small error propagation formula breaks down. Any power-law with very large or small $n$ would do this. However one might argue that these are very contrived situations. 

We can also consider
$ \Lambda_{\rm tot}=\Lambda + \Lambda_1$,
that is, two cosmological constants entering the Hamiltonian constrain, but only one contributing to the unimodular clock. This could be case of radiative corrections according to some authors~\cite{weinberg,padilla}.
In this case our argument still {\it does} go through. Although $\Lambda$ does not need to be small ($\Lambda$ and $\Lambda_1$ have opposite signs), its variance must still be small because:
\begin{equation}
    \sigma(\Lambda)=\sigma(\Lambda_{\rm tot})=\epsilon \Lambda_{{\rm tot} \,  0}\ll \Lambda_{{\rm tot} \,  0}.
\end{equation}
In other words, the variance in $\Lambda$ would have to be small with regards to the {\it total} Lambda, for the cancellation to leave $\Lambda_{\rm tot}$ with $\sigma(\Lambda_{\rm tot})\ll \Lambda_{{\rm tot} \,  0}$. 
Hence we obtain the same relation between the time of classicality $T_\star$ and the {\it total} cosmological constant, $\Lambda_{\rm tot}$. 

The same is true in the sequester model~\cite{pad,pad1}, where one removes the space-time average of the trace of the Einstein equations, so that the unimodular $\Lambda$ is not observable. Although this is true classically, one can only do this within $\sigma(\Lambda)$ in the semiclassical theory, so that $\sigma(\Lambda)$ propagates into the observable $\Lambda_{\rm obs}$ defined in~\cite{pad,pad1} (that is $ \sigma(\Lambda)=\sigma(\Lambda_{\rm obs})=\epsilon \Lambda_{{\rm obs} \,  0}\ll \Lambda_{{\rm obs} \,  0}$)\footnote{Note that if $\sigma^2(\Lambda_1)>0$, it would only tighten the constraints obtained here. We defer a full discussion to future work.}. Again, the same relation is obtained between $T_\star$ and $\Lambda_{\rm obs}$. 
The situation, however, is complicated by the fact that in~\cite{pad1} there are two clocks (a Ricci clock as well as a unimodular clock), so that a new layer comes into the argument. We defer a full analysis to a future publication. 
 


\section{In search of a superhorizon infrared scale}\label{SuperHor}

The only leeway we have in our results therefore relates to $V_c$, the comoving volume of the classical primeval patch, and obviously we would destroy our bound by letting $V_c\rightarrow \infty$, since then $\plk\rightarrow 0$ and all uncertainties would go to zero. There is, however, no reason for choosing this, quite the contrary. Indeed in this case the energy scale for classicality would be infinite, that is, at least in the minisuperspace limit the Universe would not be subjected to quantum gravity, even deep in the Planck epoch. 

Nonetheless, we use this limit as inspiration for trying to find a physically motivated context in which our bound would emerge weakened, and instead seek a context within which the Planck scale could be the scale of classicality for these theories: $\theta_\star l_P \sim 1$. 
Given that (at least for now) we cannot see beyond the last scattering surface at 13 Gpc, the scale that sets the size of the classical primeval patch only has a firm lower limit. But how big can it really be?

Current observations indicate a red spectrum of primordial scalar perturbations, implying that fluctuations become more nonlinear on large superhorizon scales. Similar to Quantum Chromodynamics (QCD), this may lead to a strong coupling scale in the deep infrared. 
Following this hypothesis, we will use the solution to the 1-loop renormalization group (RG) equation for a generic renormalizable theory (such as 4D Yang-Mills) to capture running of the scalar power spectrum:
\be
P_s (k) = \frac{A_s}{1-(n_s-1)\ln(k/k_0)},\label{power_hc}
\ee
where $n_s$ is the scalar spectral index, and $k_0 = 0.05 ~{\rm Mpc}^{-1} \simeq 221 H_0$ is the conventional pivot scale. We can see that this power spectrum diverges at:
\be
k_\star = k_0 \exp\left(1 \over n_s -1 \right).
\label{eq:kstar}
\ee
Notice that $A_s$ does not appear in this expression because it concerns a scale for a divergence, therefore erasing any fine tuning that might come from $A_s\sim 10^{-9}$. 
If we identify this scale with the (inverse of the) size of the classical primeval patch, i.e., $k_\star \sim V^{-1/3}_c = H_0/\alpha^{1/3}$ (with $\alpha$ defined in  Equation \ref{alphadef}), combining Equations (\ref{eq:theta_alpha}) and (\ref{eq:kstar}) yields:
\be
n_s = 1 -\frac{1}{32.17 + \ln(\theta_\star l_P)} \simeq 0.9689 + 1.6 \times 10^{-4} \ln(\theta_\star l_P ).
\ee
\begin{figure}
    \centering
    \includegraphics[width=\linewidth]{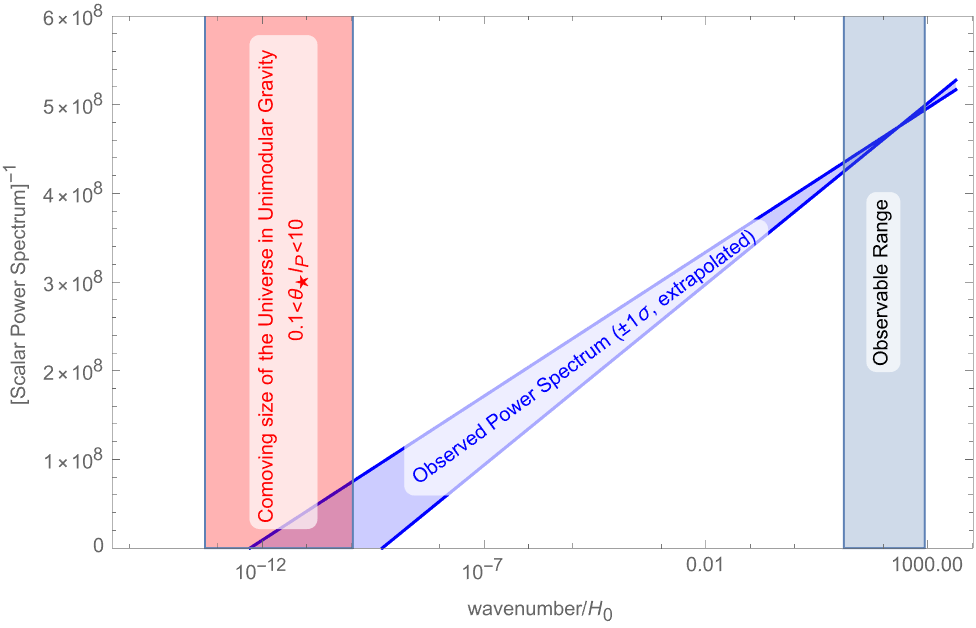}
    \caption{Observed cosmological scalar power spectrum (blue region), extrapolated to large scales ($k \ll H_0$), from the observable range (gray range). Interestingly, the wavenumber range for which $P_s^{-1}(k)$ crosses zero (i.e. scalar power spectrum diverges, Eq. \ref{eq:kstar}) coincides with the size of the comoving region that explains the observed cosmological constant in unimodular gravity (Eq. \ref{t_alpha}), assuming that a classical cosmos emerges around Planck temperature, i.e. $0.1 \lesssim \theta_\star l_P \lesssim 10$. }
    \label{fig:power}
\end{figure}
Now, imagine we require classicality to emerge somewhere within $0.1 \lesssim \theta_\star l_P \lesssim 10$. This implies:
\be
0.965 <n_s < 0.972,
\ee
which is consistent with the current combined Planck 2018 bound: $n_s =0.9665 \pm 0.0038$ \cite{Planck:2018vyg}.

This is a very surprising result, given that the two numbers, $\Lambda$ and $n_s$, are a priori not related at all. 
The only way to get the Planck scale as the scale of classicality in this model is for the classical patch to be not infinite, but about $10^{11}$ times the linear size of the current horizon, i.e. precisely the scale that becomes nonlinear given the observed scalar spectral tilt of $n_s = 0.967(4)$. This coincidence is depicted in Figure (\ref{fig:power}).

The qualitative connection between the RG flow of 3D quantum field theories and the cosmological power spectra can be made concrete in the context of McFadden \& Skenderis' Holographic Cosmology program \cite{McFadden:2009fg}. In particular, our surprising discovery of the connection between $\Lambda$ and $n_s$ through the Planck scale points to renormalizable 3D quantum theories (which have logarithimc running), and suggests that the cosmological correlations may be fully described by a finite number of couplings within these theories, which (broadly speaking) includes Chern-Simons 3-form coupling, as well as $\Phi^6$ and Yukawa couplings \footnote{K. Skenderis, private communication.}. To the best of our knowledge, determining the classes of these theories that lead to IR confinement (as suggested by our observed value of $\Lambda$), remains an open question.





\section{Conclusions}\label{conclude}
In summary, we have related the value of the cosmological constant and the energy scale for the emergence of classicality in a unimodular universe: A lower bound in the former translating into an upper bound in the latter and vice versa. The argument is robust with respect to many technicalities, with the exception of the choice of the comoving volume taken for the ``classical primeval patch'' of the Universe. If this is taken to be the current Hubble volume or even the whole of a closed Universe with a small but non-negligible $\Omega_k$ then this scale is naturally around $10^{12}$~GeV, and not the Planck scale. It would be interesting to relate this number to the amplitude of the fluctuations in inflationary models operating at this energy scale.

A second surprise found in this paper is obtained if we allow ourselves latitude for a much larger patch of classicality, $V_c$; specifically setting it to be the scale where the primordial fluctuations become divergent for a red spectrum. While this is not entirely model-independent, it suggests that a very large classical primeval patch can lift the energy scale of classicality to the Planck scale  {\it for values of $n_s$ close to the observed ones}. Note that if the scale $V_c$ were to be infinite (i.e. if an infinite Universe were to be globally classical), then the energy scale for classicality would be infinite, and not the Planck energy either. Hence it is a remarkable coincidence that the observed values of spectral tilt of $n_s = 0.967(4)$ and cosmological constant $\Lambda \simeq 7.0(2) \times 10^{-121} l^{-2}_P$ \cite{Planck:2018vyg} point to a value of $V_c$ that set the energy scale of classicality at the Planck energy. In the context of holographic cosmology \cite{McFadden:2009fg}, this coincidence suggests that the holographic 3D quantum field theory that describes our cosmological observations must be a renormalizbale theory with an IR confinement scale of $10^{11}$ times the current comoving Hubble radius.

So, what next? 

On the theoretical front a clearer understanding of what may happen as we approach the scale of the classical primeval patch is required. In the context of our first scenario with $\theta_\star \sim 10^{12}$ GeV, one may be tempted to entertain the rich zoology of eternally inflating models, but intriguingly, a positive curvature with any appreciable $|\Omega_k| > 10^{-4}$ (is believed to) rule them out entirely \cite{Kleban:2012ph}. Qualitatively, it will be hard to reconcile a near-scale invariant spectrum (as observed) with a small scale of non-linearity comparable to Hubble radius {\it in any model}. Nonetheless, interesting lessons may be learnt from 3D lattice simulations \cite{Cossu:2020yeg} in the context of super-renormalizble holographic dual theories (e.g., \cite{McFadden:2009fg,Afshordi:2016dvb,Afshordi:2017ihr}). A less-charted, but potentially more fertile territory may be a systematic study of the RG flows in renormalizable 3D quantum field theories that manifest confinement in the IR, and connects to our second scenario with $\theta_\star l_P \sim 1$. At a more foundational level, one may wonder whether there exists a holographic interpretation of the unimodular gravity. 

On the observational front, there will be ``dragons'' (or other new physics) beyond the cosmological horizon. Indeed, in the first scenario, the quantum cosmology dragons should be in our face and right around the corner.  Maybe an explanation for the infamous CMB anomalies \cite{Planck:2019evm} such as ``The axis of evil'' \cite{Land:2005ad}, ``Planck Evidence for a closed Universe'' \cite{Handley:2019tkm,DiValentino:2019qzk} (Figure \ref{fig:omegak}), or rather more subtle ``cosmological zero modes'' \cite{Afshordi:2017use} could be their tail? The fingerprints of the second scenario will be more subtle, but also more robust. For example, Equation (\ref{power_hc}) predicts the running of the spectral index to be $\frac{dn_s}{d\ln k} = (n_s-1)^2$, setting a clear target for the next generation of cosmological surveys (e.g., \cite{Li:2018epc}). Furthermore, we expect the same IR strong coupling scale, $k_\star$ (Eq. \ref{eq:kstar}) for both scalar and tensor modes. Therefore, using the same functional form as Equation (\ref{power_hc}) for tensors, we further can predict the tilt for tensors $n_t = n_s-1$ (should they ever be detected).   

The two scenarios are certainly distinguishable and falsifiable separately. For example, the observation of topological defects left over from a phase transition at an energy scale above $10^{12}$~GeV would kill the first scenario (but we are not holding our breath). Furthermore, given that such low scale of classicality only allows for low-scale inflation, then inflationary modes should be unobservable\footnote{We thank Tony Padilla for pointing this out.}.  
There are also other intriguing but very model-dependent possibilities for the first scenario, for example regarding the amplitude of scalar fluctuations. The ratio between $\theta_\star$ and the Planck scale is then $10^{-7}$, not too different from $10^{-5}$. Could the value of $\Lambda$ and that the amplitude of the primordial fluctuations be related?

We close with general comments regarding other work on the cosmological constant. We note that the argument we presented here gives a (minimal) width for the distribution of $\Lambda$, and not its central value, solving what Weinberg referred to as the ``new cosmological constant problem'' \cite{Weinberg:2000yb}. The ``old cosmological problem'' of why $\Lambda =0$ is contained within (or at the centre) of this distribution may find a solution through the non-perturbative structure of quantum gravity (e.g., \cite{Hawking:1984hk,Afshordi:2008xu,Aslanbeigi:2011si,Kaloper:2022jpv}). We note also that the scale $V_c$ could appear in the variance $\sigma(\Lambda)$ itself, as happens in causal set models~\cite{sorkin-review,Ahmed:2002mj,Zwane:2017xbg}, where $\Lambda$ is a Poissonian process. This does not affect our argument, since the point made here is that $\sigma(\Lambda)$ must be smaller than the central value $\Lambda_0$ on whatever scale of classicality, $V_c$, we have defined. 
\begin{acknowledgments}
We thank Bruno Alexandre, Davide Gaiotto, Tony Padilla, and Kostas Skenderis for discussions related to this paper.  NA is funded by the University of Waterloo, the National Science and Engineering Research Council of Canada (NSERC) and the Perimeter Institute for Theoretical Physics.
Research at Perimeter Institute is supported by the Government of Canada through Industry Canada and by the Province of Ontario through the Ministry of Economic Development \& Innovation. JM is funded by the STFC Consolidated Grant ST/T000791/1. JM also  thanks the Perimeter Institute for hospitality and support.
\end{acknowledgments}

\bibliography{Ref.bib}

\end{document}